\documentclass[onecolumn,preprintnumbers,amsmath,amssymb,prl,superscriptaddress,endfloats*]{revtex4}

\usepackage{graphicx}
\usepackage{dcolumn}
\usepackage{bm}

\begin{document}

\title{Experimental Violation of Bell's Inequality in Spatial-Parity Space}
\author{Timothy Yarnall}
\affiliation{Quantum Imaging Laboratory, Departments of Electrical
\& Computer Engineering and Physics, Boston University, Boston,
Massachusetts 02215-2421, USA} \homepage{http://www.bu.edu/qil}
\author{Ayman F. Abouraddy}
\affiliation{Research Laboratory of Electronics, Massachusetts Institute of Technology,
Cambridge, Massachusetts 02139-4307, USA}
\author{Bahaa E. A. Saleh}
\author{Malvin C. Teich}
\affiliation{Quantum Imaging Laboratory, Departments of Electrical \& Computer
Engineering and Physics, Boston University , Boston, Massachusetts 02215-2421, USA}
\homepage{http://www.bu.edu/qil}
\date{\today}

\begin{abstract}
We report the first experimental violation of Bell's inequality in
the spatial domain using the Einstein--Podolsky--Rosen state.
Two-photon states generated via optical spontaneous parametric
downconversion are shown to be entangled in the parity of their
one-dimensional transverse spatial profile.  Superpositions of Bell
states are prepared by manipulation of the optical pump's transverse
spatial parity---a classical parameter. The Bell-operator
measurements are made possible by devising simple optical
arrangements that perform rotations in the one-dimensional
spatial-parity space of each photon of an entangled pair and
projective measurements onto a basis of even--odd functions. A
Bell-operator value of $2.389 \pm 0.016$ is recorded, a violation of
the inequality by more than 24 standard deviations.
\end{abstract}
\maketitle

\textit{Introduction}---Although Einstein, Podolsky, and Rosen (EPR)
\cite{Einstein35PR} couched their challenge to the completeness of
quantum mechanics in the language of continuous spatial parameters
of a two-particle quantum state, most of the subsequent work
investigating their claim relied on discrete degrees of freedom
\cite{Bouwmeester00Book}. In particular, Bell \cite{Bell64}
described an approach to delineate quantum theory from those that
subscribe to local realism in terms of correlations between two
spin-$\frac{1}{2}$ particles. Surprisingly, after more than seventy
years of studying the EPR state, the quantum nonlocality potentially
exhibited by it has \textit{not} been experimentally demonstrated
\textit{in the spatial domain}.

While the original paradox has been realized using the EPR state in
numerous experiments \cite{Ou92PRL}, including one in the spatial
domain \cite{Howell04PRL2}, it was posited by Bell himself
\cite{Bell86} that the EPR state should not violate a Bell-type
inequality, and hence does not violate local realism, since its
associated Wigner distribution \cite{Wigner32PR} is positive
everywhere. It has since been recognized that this statement is not
correct \cite{Johansen97PLA}. The challenge one faces to demonstrate
a Bell-inequality violation in the spatial domain is to identify
operators that perform rotations and projective measurements on
suitably defined observables of the infinite-dimensional Hilbert
space associated with the spatial profile of entangled photon pairs.
A recent proposal to achieve this \cite{Oemrawsingh04PRL}, relying
on orbital angular momentum observables manipulated by spiral phase
plates, was subsequently retracted \cite{Oemrawsingh05PRL}. A
modified proposal using the same variables \cite{AielloPRA05} relies
on \textit{projecting} the EPR state in the spatial domain onto a
two-dimensional \textit{subspace} and has thus far not been
demonstrated experimentally; nor have other proposals
\cite{Larsson:022102}. Another approach that retains the entire
Hilbert space relies on a set of `pseudospin' operators that have
been recently constructed \cite{Chen02PRL} and that provide a bridge
between states described by continuous and discrete variables. In
essence, these proposed operators map the infinite-dimensional
Hilbert space of the EPR state onto a smaller-dimensional space
\cite{Brukner03PRA}. Nevertheless, the physical realization of these
operators presents daunting experimental difficulties
\cite{Chen02PRL,Ralph00PRL}.

In this Letter, we present a conclusive experimental violation of
Bell's inequality \textit{in the spatial domain}, and thus
demonstrate quantum nonlocality using the EPR state produced by
spontaneous parametric downconversion (SPDC) \cite{Harris67PRL}.
This is made possible by constructing photonic `pseudospin'
operators in the spatial domain using an approach that we recently
described \cite{Abouraddy07PRA}. These operators make use of the
spatial parity (even--odd) of the one-dimensional (1D) transverse
field of single photons. This entails mapping the
infinite-dimensional Hilbert space of the photon's transverse
spatial profile onto a space of dimension two. It is important to
note that the potentially infinite-dimensional Hilbert space of
transverse modes (limited by the effective numerical aperture of the
experimental arrangement) is \textit{mapped} onto a two-dimensional
space \cite{Brukner03PRA} and is \textit{neither truncated} as in
Ref.~\cite{Mair01Nat} \textit{nor projected}  onto a
smaller-dimensional subspace as in Ref.~\cite{AielloPRA05}.

\textit{Spatial-Parity Space}---To introduce the above-mentioned
mapping, consider a one-photon state in the spatial domain
$|\Psi\rangle=\int dx\psi(x)|1_{x}\rangle$, where $\int
dx|\psi(x)|^{2}=1$. This state is potentially of infinite
dimensionality as governed by the expansion into an orthonormal
functional basis $\psi(x)=\sum_{n}c_{n}\xi_{n}(x)$, such that
$|\Psi\rangle=\sum_{n}c_{n}|n\rangle$, where $|n\rangle=\int
dx\xi_{n}(x)|1_{x}\rangle$. We map this state onto two `levels' of a
qubit: the even $\psi_\textrm{e}(x)=\frac{1}{2}\{\psi(x)+\psi(-x)\}$
and odd $\psi_\textrm{o}(x)=\frac{1}{2}\{\psi(x)-\psi(-x)\}$
components of the photon transverse distribution, which are
orthogonal $\int dx \psi_\textrm{e}^*(x)\psi_\textrm{o}(x) = 0$. The
one-photon state may be recast in the even $|\textrm{e}\rangle$ and
odd $|\textrm{o}\rangle$ basis of this 2D space of spatial parity,
$|\Psi\rangle=\alpha|\textrm{e}\rangle+\beta|\textrm{o}\rangle$,
where $\alpha|\textrm{e}\rangle=\sum_{n}c_{2n}|2n\rangle$ and
$\beta|\textrm{o}\rangle=\sum_{n}c_{2n+1}|2n+1\rangle$. The
fundamental insight provided in this paper is the isomorphism
between the `pseudospin' approach and the that of spatial-parity
space.

It has been shown \cite{Abouraddy07PRA} that the `pseudospin'
operators corresponding to the well-known Pauli operators, as well
as other relevant operators (such as rotation and projection
operators), are easily implementable on this new Hilbert space. The
two most relevant operators for this Letter are the parity rotator
and the parity analyzer. The parity rotator rotates the state
function in parity space, just as a polarization rotator rotates the
polarization vector. This operator is implemented using a phase
plate that introduces a phase $\theta$ between the $x\geq 0$ and
$x<0$ half planes (i.e., having transmissivity
$e^{i\frac{\theta}{2}\textrm{sgn}(x)}$), corresponding to a rotation
$\theta$ on a Poincar{\'e} sphere \cite{Abouraddy07PRA}. The parity
analyzer is a device that separates the even and odd components of
the state into two separate spatial paths, thus projecting the state
onto the even--odd basis. Our implementation of a parity analyzer
uses a Mach--Zehnder interferometer (MZI) at zero relative path
length delay with a spatial flipper (viz., a device that produces a
mirror image of the incident field,
$\varphi(x)\rightarrow\varphi(-x)$, implemented in our case by a
mirror) inserted in one arm. The resulting interferometer becomes
parity sensitive (PS-MZI) \cite{Abouraddy07PRA,Sasada03PRA}.

\begin{figure}[t]
  \includegraphics[width=3in]{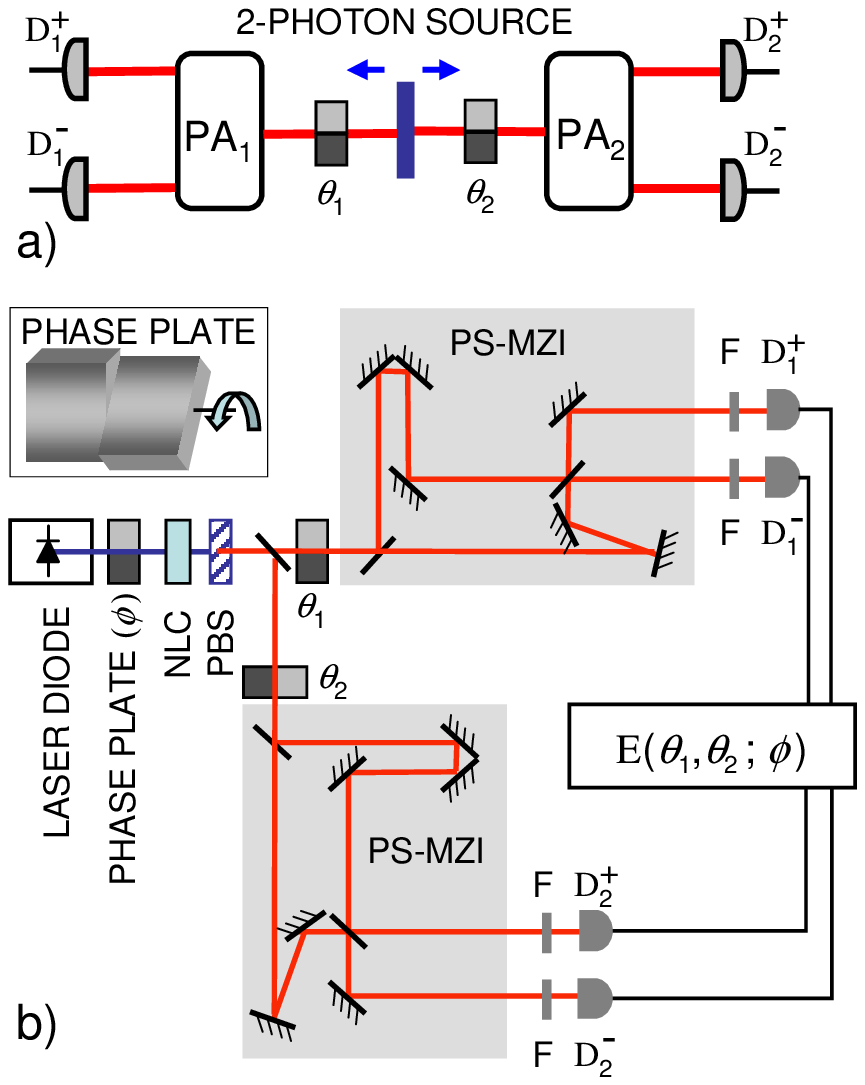}\\
  \caption{(Color online) (a) Notional configuration for testing Bell's inequality. PA: parity analyzer;
  $\theta$: parity rotator; D: detector. (b) Schematic of the experimental arrangement. NLC: nonlinear optical
  crystal; PBS: polarizing beam splitter; PS-MZI: parity sensitive Mach--Zehnder interferometer;
  F: interference filter; D avalanche photodetector.  The inset depicts the construction of a phase plate (parity rotator)
  used at three locations in the setup: after the laser diode ($\phi$) and immediately preceding each PS-MZI ($\theta_{1}$ and $\theta_{2}$).  It comprises two
  glass microscope slides, abutted at the origin of the transverse dimension $x=0$ that can be tilted with respect
  to each other, thus introducing a relative phase between the two halves of the plane.}
  \label{setup}
\end{figure}

\textit{Experimental Arrangement}--- A generic configuration for the
experimental violation of Bell's inequality is shown in Fig.
\ref{setup}(a). A two-photon source directs each photon to an SO(2)
rotation operator (characterized by angular settings $\theta_{1}$
and $\theta_{2}$) followed by a projective measurement. Coincidence
measurements between the detectors $\textrm{D}_1$ and $\textrm{D}_2$
for various settings of $\theta_1$ and $\theta_2$ are used to obtain
the correlations between a pair of dichotomic outcomes needed for
assessing a violation of Bell's inequality in the CHSH formulation
\cite{Clauser69PRL}. In our conception, the SO(2) operator is a
parity rotator and the projective measurements on each photon are
performed using parity analyzers. Our experimental arrangement is
shown schematically in Fig. \ref{setup}(b). Light from a linearly
polarized laser diode (center wavelength 405 nm, power 50 mW)
emitting in an even-symmetry spatial mode $\psi_{\textrm{even}}(x)$
is passed through a phase plate that serves as a parity rotator
\cite{Abouraddy07PRA}. The angle $\phi$ determines the spatial
parity of the exiting pump beam, which illuminates a 1.5-mm-thick
$\beta$-barium borate (BBO) nonlinear optical crystal (NLC) in a
collinear type-I configuration (signal and idler photons have the
same polarization, orthogonal to that of the pump). The collinear
signal and idler photons are separated by a beam splitter; each of
the exiting photons passes through a parity rotator, set at
$\theta_1$ and $\theta_2$. Each output photon then enters a parity
analyzer and then exits to the detectors. The unconverted pump light
is removed with the help of a polarizing beam splitter (PBS) placed
after the NLC as well as by interference filters F (centered at 810
nm, 10-nm bandwidth) in front of the
\textit{multimode}-fiber-coupled detectors
$\textrm{D}_1^+,\,\textrm{D}_1^-,\,\textrm{D}_2^+\,$ and
$\textrm{D}_2^-$ (EG\&G SPCM-AQR-15-FC). It is imprtant to note that
we do not couple the photons into single-mode fibers as is the usual
practice in projecting orbital angular momentum states onto a single
spatial mode \cite{Mair01Nat,Oemrawsingh05PRL}; as mentioned above,
we aim to collect \textit{all} the available spatial modes. The
outputs of these detectors are fed to coincidence circuits and
thence to counters, from which a correlation function
E$(\theta_1,\,\theta_2;\,\phi)$ (to be defined shortly) is obtained.

\textit{Quantum-State Preparation}---We produce the different
two-photon states investigated here by manipulating the spatial
parity of the pump profile $E_{p}(x)$, a classical parameter, via
its passage through a phase plate that produces the classical field
distribution
\begin{equation}\label{Pump}
E_{p}(x)=\cos(\phi/2)\ \psi_{\textrm{even}}(x)+i\,\sin(\phi/2) \
\psi_{\textrm{odd}}(x),
\end{equation}
where $\phi$ is the angle of parity rotation imparted by the phase
plate to the pump, and $\psi_{\textrm{odd}}(x)=\textrm{sgn}(x)\,
\psi_{\textrm{even}}(x)$. The two-photon quantum state produced in
1D by SPDC \cite{Harris67PRL,Abouraddy07PRA} is
$|\Psi\rangle=\int\int dxdx'\psi(x,x')|1_{x},1_{x'}\rangle$, where
\begin{equation}\label{Schmidt}
\psi(x,x')=E_{p}(\frac{x+x'}{2})\xi(\frac{x-x'}{2})=\sum_{n=0}^{N}\lambda_{n}\phi_{n}(x)\zeta_{n}(x'),
\end{equation}
where $\xi(x)$ is a function of width much smaller than that of
$E_{p}(x)$, representing the correlation between the emission
locations of the two photons, $\sum_{n=0}^{N}\lambda_{n}^{2}=1$, and
$\{\phi_{n}(x)\}$ and $\{\zeta_{n}(x)\}$ are two orthonormal sets of
functions in the Schmidt decomposition \cite{law:127903}.

One approach to estimating the number of transverse modes $N$ in the
state function's Schmidt decomposition is to approximate $E_{p}(x)$
and $\xi(x)$ by Gaussian functions with appropriate widths, in which
case an analytical expression for $N$ exists \cite{law:127903},
$N=\frac{1}{4}(w/\sqrt{\lambda_{p}\ell/8}+\sqrt{\lambda_{p}\ell/8}/w)^2\approx4000$,
where the pump width $w=1.1$ mm, the NLC thickness $\ell=1.5$ mm,
and the pump wavelength $\lambda_{p}=405$ nm. One may also obtain
the Schmidt decomposition directly by diagonalizing the function
$\eta(x_{1},x_{2})=\int dx'\psi^{*}(x_{1},x')\psi(x_{2},x')$ and
counting modes with eigenvalues larger than some threshold, say
$\lambda_{n}>0.01\lambda_{\textrm{max}}$, which yields a value of
$N\approx3500$. Thus the state that we use inhabits a Hilbert space
of extremely high dimensionality $N\times N$.

The even $\psi_{\textrm{even}}(x)$ and odd $\psi_{\textrm{odd}}(x)$
components of the pump spatial field distribution generate through
the process of SPDC the two-photon spatial-parity states
$|\Phi^+\rangle$ and $|\Psi^+\rangle$, respectively, where
$|\Phi^+\rangle =
\frac{1}{\sqrt{2}}\left\{|\textrm{e}\rangle_1|\textrm{e}\rangle_2+|\textrm{o}\rangle_1|\textrm{o}\rangle_2\right\}$
and $ |\Psi^+\rangle =
\frac{1}{\sqrt{2}}\left\{|\textrm{e}\rangle_1|\textrm{o}\rangle_2+|\textrm{o}\rangle_1|\textrm{e}\rangle_2\right\}$,
as detailed in Ref. \cite{Abouraddy07PRA}. This may be understood
intuitively by noting that an even-parity (odd-parity) pump results
in a Schmidt decomposition of the two-photon state function in Eq.~(
\ref{Schmidt}) composed of a sequence of products of only even--even
or odd--odd (even--odd or odd--even) functions. The two-photon state
generated by the pump distribution in Eq.~(\ref{Pump}) is thus
\begin{equation}\label{MaximallyEntangledState}
|\Psi\rangle=\cos(\phi/2)\,|\Phi^+\rangle+i\,\sin(\phi/2)\,|\Psi^+\rangle.
\end{equation}
It is important to note that this two-photon state is
\emph{maximally entangled} regardless of the value of $\phi$ (the
concurrence \cite{Wootters97PRL} of the state is unity,
independent of $\phi$) and thus allows for a maximal violation of Bell's inequality.

For the purposes of this Letter only three values of $\phi$ are
considered, namely $\phi=0$ (corresponding to an even pump),
$\frac{\pi}{2}$ (pump in an equal superposition of even and odd
components), and $\pi$ (odd pump), leading to the following
two-photon states: $|\Phi^+\rangle$,
$\frac{1}{\sqrt{2}}\{|\Phi^+\rangle+i\,|\Psi^+\rangle\}$ and
$|\Psi^+\rangle$, respectively.  A fourth state,
$\frac{1}{\sqrt{2}}\{|\Phi^+\rangle+|\Psi^+\rangle\}$ (i.e., without
the factor $i$) is prepared using a pump having the distribution
$\frac{1}{\sqrt{2}}\{\psi_{\textrm{even}}(x)+\psi_{\textrm{odd}}(x)\}$.
This is achieved by removing the phase plate used to control the
pump parity (Fig.~\ref{setup}) and replacing it with an opaque
screen that blocks the field on the positive $x$ axis. The resulting
two-photon state
$\frac{1}{\sqrt{2}}\{|\Phi^+\rangle+|\Psi^+\rangle\}=\frac{1}{\sqrt{2}}\{|\textrm{e}\rangle_1+|\textrm{o}\rangle_1\}
\otimes
\frac{1}{\sqrt{2}}\{|\textrm{e}\rangle_2+|\textrm{o}\rangle_2\}$ is
separable and as such will not violate Bell's inequality.

\textit{Bell-Operator Measurements}---After their preparation, each
photon in the two-photon states undergoes a rotation in parity space
followed by a projective measurement implemented by a phase plate
and a PS-MZI, respectively. Photon detection is done in coincidence
and the value $+1 \ (-1)$ is associated  with even (odd) outcomes.
One can then estimate the correlation function
E$(\theta_1,\theta_2;\phi)$ between the \textit{parities} of the two photons
as a function of their respective rotations. It should be noted that
the rotation employed here is about an orthogonal axis on the
Poincar\'{e} sphere to the one often reported in polarization-based
experiments. This distinction becomes important when considering
tests of quantum physics more stringent than that of Bell
\cite{Leggett03FoundPhys}.

Evaluation of the Bell operator in the CHSH form requires
measurement of four correlations requiring two settings for each
parity rotator
\begin{equation}
\mathcal{B}=\left|\textrm{E}(\theta_1,\theta_2)+\textrm{E}(\theta_1,\theta_2')+\textrm{E}(\theta_1',\theta_2)-\textrm{E}(\theta_1',\theta_2')\right|
\leqslant 2,
\end{equation}
where we omit the implicit dependence on $\phi$ for simplicity.
Quantum theory predicts the correlation function for the maximally
entangled state given in Eq.~(\ref{MaximallyEntangledState}) to be
E$(\theta_1,\theta_2 ; \phi) = \cos(\theta_1+\theta_2+\phi)$.
Accordingly, $\mathcal{B}$ attains the maximum value of $2\sqrt{2}$
when $\theta_1=\theta_2=\frac{\pi}{8}-\frac{\phi}{2}$ and
$\theta_1'=\theta_2'=\frac{13\pi}{8}-\frac{\phi}{2}$. When the sate
is separable, as is the case for
$\frac{1}{\sqrt{2}}\{|\Phi^+\rangle+|\Psi^+\rangle\}$, the predicted
correlation function is E$(\theta_1,\theta_2) = 0$. The lack  of
dependence on the setting of either parity rotator can be understood
by noting the axis of rotation on a Poincar\'{e} sphere. In this
case each photon's state of parity is a point that lies \textit{on}
the axis of rotation chosen in our experimental configuration and
the correlation function is thereby unchanged.

\begin{figure}
  \includegraphics[width=3.4in]{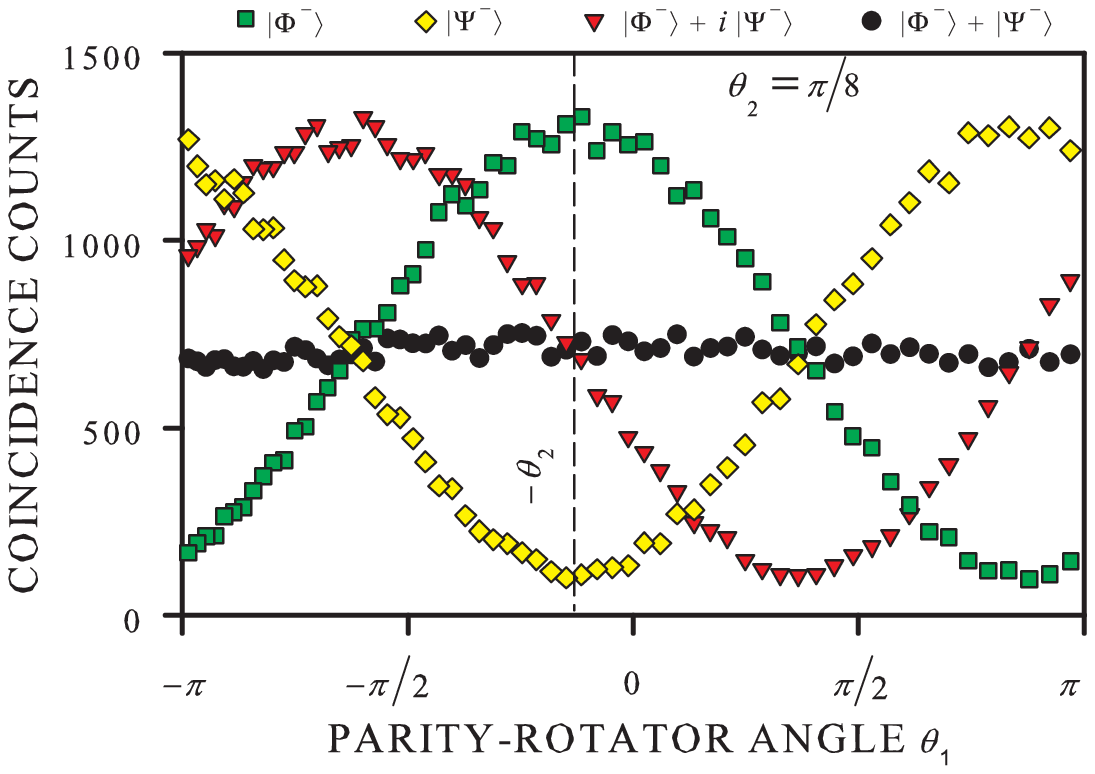}\\
  \caption{(Color online) Coincidence counts between
  $\textrm{D}_1^+$ and $\textrm{D}_2^+$
  as a function of $\theta_1$ for fixed $\theta_2=\frac{\pi}{8}$ and three values of
  $\phi$:
  the green $\blacksquare$ for $\phi=0$ (even pump), the red $\blacktriangledown$ for $\phi=\frac{\pi}{2}$ (pump in equal superposition of even and odd), and the yellow $\blacklozenge$ for
  $\phi=\pi$ (odd pump).  The black $\bullet$ represents the same measurement when the initial phase plate is replaced by an
  opaque screen that blocks the positive $x$ axis, which leads to a separable two-photon state. A side-effect of the screen is to
  halve the number of photons produced via SPDC; for purposes of comparison with the previous cases we therefore plotted twice the
  recorded values in the figure.}
  \label{slice}
\end{figure}

\begin{figure*}[b]
  \includegraphics[width=7in]{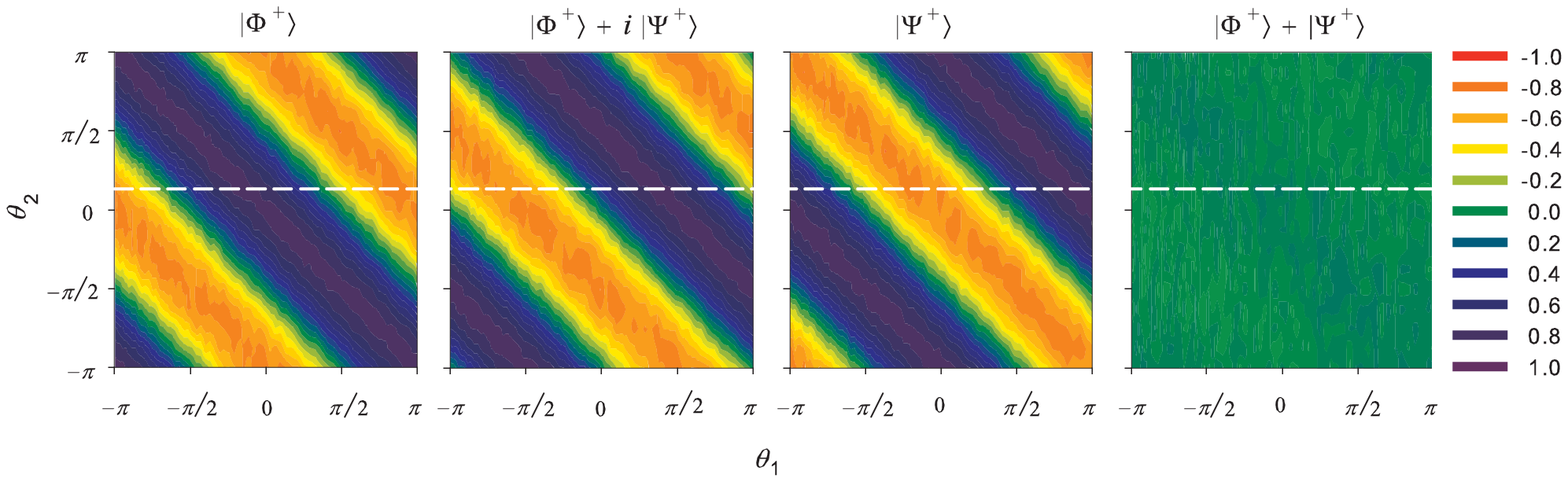}\\
  \caption{(Color online) The correlation function E$(\theta_1,\theta_2;\,\phi)$ derived from
  coincidence measurements for three different maximally entangled two-photon states, corresponding to different settings of the pump parity angle $\phi$ (from left to
  right $\phi=0$, $\phi=\frac{\pi}{2}$, and $\phi=\pi$).  The final
  (right) panel shows E$(\theta_1,\theta_2)$ for a separable state.  The dashed white line indicates the cross-section of the data presented in Fig.~\ref{slice}.
  }
  \label{correlation}
\end{figure*}

\textit{Experimental Results}---Although evaluation of the Bell
operator requires that E$(\theta_1,\theta_2;\phi)$  be measured at
only four points in the $(\theta_1,\theta_2)$ plane; measuring E
over the full range of values for $\theta_1$ and $\theta_2$ provides
much more information and offers insight into the effect of the pump
parity on the Bell operator. Measurements of the correlation
function, for each of the four states prepared, were performed by
varying $\theta_1$ and $\theta_{2}$ over $2\pi$ radians.
Figure~\ref{slice} provides an example of the coincidence rate
recorded by the `even' detectors after the parity analyzers when
$\theta_2=\frac{\pi}{8}$; the visibility of the recorded coincidence
sinusoids for each of the maximally entangled states is
$\approx84\%$.  This lower-than-expected visibility is a result of
imperfect alignment of the PS-MZIs. The full landscapes for
E$(\theta_{1},\theta_{2};\phi)$ are shown in Fig.~\ref{correlation}.
The functional form of E$(\theta_{1},\theta_{2};\phi)$ for the
entangled states are clearly seen to be 2D sinusoids in
$\theta_{1}+\theta_{2}$ with the fringes shifted according to the
value of $\phi$. The separable state reveals a flat correlation
landscape. The measured values of the Bell operator for each state
are presented in Table~\ref{table1}. All three entangled states
demonstrate clear violations of Bell's inequality.
\begin{table}
\caption{\label{table1}Measured values of the Bell operator for the
four two-photon states investigated, and the amount of the violation
of Bell's inequality.}
\begin{ruledtabular}
\begin{tabular}{ccc}
State & Bell Operator $\mathcal{B}$&Violation\\
\hline
$|\Phi^+\rangle$ & $2.389 \pm 0.016$ & $24.2\sigma$\\
$|\Psi^+\rangle$ & $2.375 \pm 0.018$ & $20.5\sigma$\\
$|\Phi^+\rangle+i\,|\Psi^+\rangle$  & $2.373 \pm 0.019$ & $19.7\sigma$\\
$|\Phi^+\rangle+\,\,\,|\Psi^+\rangle$ & $0.057 \pm 0.017$ & None
expected
\end{tabular}
\end{ruledtabular}
\end{table}

\textit{Conclusion}---We have shown that the entangled state
proposed in the original embodiment of the EPR paradox violates a
Bell inequality in the spatial domain, and thus is at odds with any
local-hidden-variables theory.  We identified dichotomic observables
based on the spatial parity of each particle's transverse spatial
distribution. This identification, combined with our demonstrated
abilities to perform rotations in parity space, and projections onto
an even--odd basis, enabled us to carry out a direct test of Bell's
inequality in the spatial domain.  The results demonstrate clear
violations of the bounds imposed by local realism, in agreement with
quantum theory.

\begin{acknowledgments}\textit{Acknowledgments}---This work was supported by a U.~S.~Army
Research Office (ARO) Multidisciplinary University Research
Initiative (MURI) Grant and by the Center for Subsurface Sensing and
Imaging Systems (CenSSIS), an NSF Engineering Research Center. This
work is sponsored by the National Aeronautics and Space
Administration under Air Force Contract \#FA8721-05-C-0002.
Opinions, interpretations, recommendations and conclusions are those
of the authors and are not necessarily endorsed by the United States
Government. A.F.A. acknowledges the generous support and
encouragement of Y.~Fink and J.~D.~Joannopoulos.
\end{acknowledgments}


\end{document}